\documentclass[twocolumn,aps,prb,showkeys, 10pt]{revtex4-2}

\makeatletter
\def\@fnsymbol#1{%
  \ensuremath{%
    \ifcase#1\or
      *\or
      **\or
      \dagger\or
      \ddagger\or
      \mathsection\or
      \mathparagraph
    \else
      \@ctrerr
    \fi}}
\makeatother

\usepackage{chemformula}
\usepackage{graphicx}
\usepackage{amsfonts}
\usepackage{algorithm}
\usepackage{algpseudocode}
\usepackage{subcaption}
\usepackage{float}
\usepackage{gensymb}
\usepackage[inkscapelatex=false]{svg}
\usepackage{dcolumn}
\usepackage{bm}
\usepackage[version=4]{mhchem}
\usepackage{booktabs}
\usepackage[hidelinks]{hyperref}

\begin{document}

\raggedbottom

\title{Quantitatively Probing the Tilt Angle Failures in 4D-STEM Polarisation Analysis}

\author{Fintan G. Hardy}
\author{Noah Schnitzer}
\author{Mariana Palos}
\author{Geri Topore}
\affiliation{Department of Materials, London Centre of Nanotechnology, Imperial Henry Royce Institute, Imperial College London, SW7 2AZ, UK}

\author{Aron Walsh}
\email{a.walsh@imperial.ac.uk}
\affiliation{Department of Materials, London Centre of Nanotechnology, Imperial Henry Royce Institute, Imperial College London, SW7 2AZ, UK}

\author{Michele Shelly Conroy}
\email{mconroy@imperial.ac.uk}
\affiliation{Department of Materials, London Centre of Nanotechnology, Imperial Henry Royce Institute, Imperial College London, SW7 2AZ, UK}

\begin{abstract}

Four-dimensional scanning transmission electron microscopy (4D-STEM) is increasingly used to map polarisation in ferroic textures such as domain walls, vortices, merons and skyrmions. These textures are accompanied by local strain, lattice rotation and misorientation, which also modulate the diffraction features used to infer polar order. Even small deviations from the zone axis can therefore produce contrast that is indistinguishable from a polarisation signal. Here we quantify the tilt sensitivity of two polarisation-sensitive signals, diffraction disc intensity and Kikuchi band intensity, in three ferroelectrics with distinct polar distortions: PbTiO$_3$, BaTiO$_3$ and Cu$_3$B$_7$O$_{13}$Cl boracite. We find that tilts of only 0.01--0.05\degree reverse most Friedel-pair disc asymmetries, and even the most robust disc pairs reverse by 0.04\degree in BaTiO$_3$, 0.14\degree in PbTiO$_3$ and 0.18\degree in boracite, producing contrast equal and opposite to that of the polarisation. Kikuchi band contrast is the only signal that reports absolute polarity without knowledge of the specimen thickness, but it does so only up to 0.09\degree in PbTiO$_3$, 0.015\degree in BaTiO$_3$ and 0.02\degree in boracite. From these results we define tilt tolerances for each signal and material, and propose practical criteria for distinguishing polarisation contrast from misorientation. The analysis also applies to scanning electron microscopy-based diffraction, including electron backscatter diffraction.

\end{abstract}

\maketitle


\section{\label{sec:level1}Introduction}

Local polarisation structures in ferroelectric materials underpin a wide range of functional behaviours, from domain switching processes to the formation of complex topological states such as vortices and skyrmions \cite{paruchNanoscaleStudiesDomain2006, meier2022ferroelectric, dasObservationRoomtemperaturePolar2019, nahasDiscoveryStableSkyrmionic2015, yadavObservationPolarVortices2016, hadjimichaelMetalFerroelectricSupercrystals2021a, mooreChargedDomainWall2022}. Advances in engineering these polar textures have increased the need for techniques capable of mapping polarisation with high spatial resolution and sensitivity.

Piezoresponse force microscopy (PFM) has been the principal technique for mapping ferroelectric domains at the nanoscale, providing spatially resolved measurements of the local electromechanical response and enabling studies of domain structures, polarisation switching and domain wall dynamics. Advances in PFM have played an important role in understanding how ferroelectric domains form, evolve and respond to applied electric fields \cite{Gruverman2019PFM, Gruverman2006Nanoscale, McQuaid2011FluxClosure, McQuaid2017Boracite, Jungk2007PFM, checaOndemandNanoengineeringInplane2025}. However, as interest has shifted towards increasingly complex polar textures and local variations in polarisation associated with defects, interfaces and domain walls, there is a growing need for complementary characterisation techniques. PFM primarily probes the near-surface electromechanical response, and quantitative interpretation can be complicated by electrostatic contributions and other non-ferroelectric signals. Alternative approaches, including X-ray diffraction and optical second-harmonic generation, provide information on polar order but often lack the spatial resolution needed to resolve individual domain walls and local variations in polarisation \cite{yefanovMappingContinuousReciprocal2014, zhangCharacterizationDomainDistributions2018}. Electron microscopy and diffraction offer a complementary approach, combining sensitivity to local crystal structure with high spatial resolution. This makes them particularly useful for investigating ferroelectric domain walls, dislocations, interfaces and complex polar topologies, where polarisation must be considered alongside local strain, lattice rotations and other structural variations \cite{mooreAberrationCorrectedSTEM2021}.

Several electron diffraction-based techniques have been developed to extract quantitative polarisation information from convergent beam electron diffraction (CBED) patterns. In polar crystals, the loss of inversion symmetry modifies the crystal electrostatic potential and the phases of the structure factors. Through dynamical electron scattering, interference between diffracted beams can produce measurable intensity differences between inversion-related reflections, resulting in a breakdown of Friedel symmetry \cite{shaoEmergentChiralityPolar2023}. These intensity asymmetries provide a means of determining local polarisation from CBED patterns. Momentum transfer to the electron beam has been used to map local polarisation and investigate the toroidal and chiral order of polar vortices through measurements of orbital angular momentum transfer \cite{nguyenTransferringOrbitalAngular2023b}. Other approaches quantify polarisation-related asymmetries through the normalised intensity difference between Friedel-related Bragg discs \cite{susarlaEmergenceThreedimensionalChiral2023}, or analyse changes in the scattered intensity distribution using centre-of-mass measurements \cite{hoglundTopologyDrivenVibrationsChiral}.

An alternative approach uses intensity asymmetries within Kikuchi bands, which arise from thermally diffusely scattered electrons undergoing subsequent diffraction within the crystal. The intensity distribution within these bands contains contributions that depend on the phases of the crystal structure factors. In non-centrosymmetric crystals, these contributions can produce asymmetric Kikuchi-band intensities even under symmetric scattering conditions, providing a means of determining crystal polarity \cite{BirdWright1989}. The original theoretical treatment also demonstrated that these contributions depend on specimen thickness, with the polarity-sensitive term being particularly relevant for relatively thin crystals. Kikuchi-band intensity asymmetries have subsequently been used to map the in-plane polarisation of ferroelectric skyrmions and merons, with reduced sensitivity to local crystal mis-tilt compared with Bragg-disc intensities reported under the conditions investigated \cite{shaoEmergentChiralityPolar2023}. More recently, centre-of-mass shifts in Kikuchi-band intensities have been used to map nanoscale polar textures in quantum paraelectric \ce{SrTiO3} at helium temperatures, demonstrating the sensitivity of this approach to weak local polar order \cite{zhangNanoscalePolarLandscapes2025}.

Other diffraction-based methods include applying machine learning to the full diffraction pattern \cite{ludackaImagingStructureAnalysis2024c} and extracting polar displacements from the imaginary component of the exit-wave cepstrum \cite{holtzMappingPolarDistortions2023b}. Although approaches have also been developed to recover the polarisation component along the electron beam direction using deep learning \cite{byunImaging3DPolarizationDynamics2025}, the intensity-based methods examined in this study primarily probe the polarisation projected perpendicular to the incident beam, providing a two-dimensional representation of the polarisation state.

In practice, however, the assumption of perfect zone-axis alignment is rarely possible. While specimen misalignment may arise from microscope or sample preparation limitations, local crystallographic rotations are also an intrinsic feature of many ferroelectric microstructures. Domain walls, dislocations, strain gradients, and higher-order polar topologies can all generate local variations in crystal orientation, resulting in spatially varying specimen tilt even within a nominally aligned sample. In complex domain configurations, such as flux-closure structures and polar vortices, neighbouring domains are often connected through continuous lattice rotations or crystallographic misorientations. As a result, it is generally impossible to orient all domains simultaneously along a common zone axis. For example, in tetragonal \ce{PbTiO3}, neighbouring ferroelastic domains may differ in orientation by approximately 90\degree \cite{tovaglieriInvestigatingDomainStructures2025, tangAtomicscaleMappingDipole2014, tangObservationPeriodicArray2015}, meaning that optimisation of diffraction conditions for one domain inevitably introduces a degree of misalignment in adjacent regions. The importance of accounting for local crystal orientation has also been highlighted by a reciprocity-based 4D-STEM approach that uses Kikuchi-band information to determine local misorientation and investigate ionic displacements across inclined ferroelectric domains \cite{Dushimineza2026}. Consequently, tilt-induced diffraction effects are not merely an experimental artefact but are fundamentally coupled to many of the ferroelectric phenomena that these techniques are intended to investigate.

\begin{figure}
    \centering
    \includegraphics[width=8.3cm]{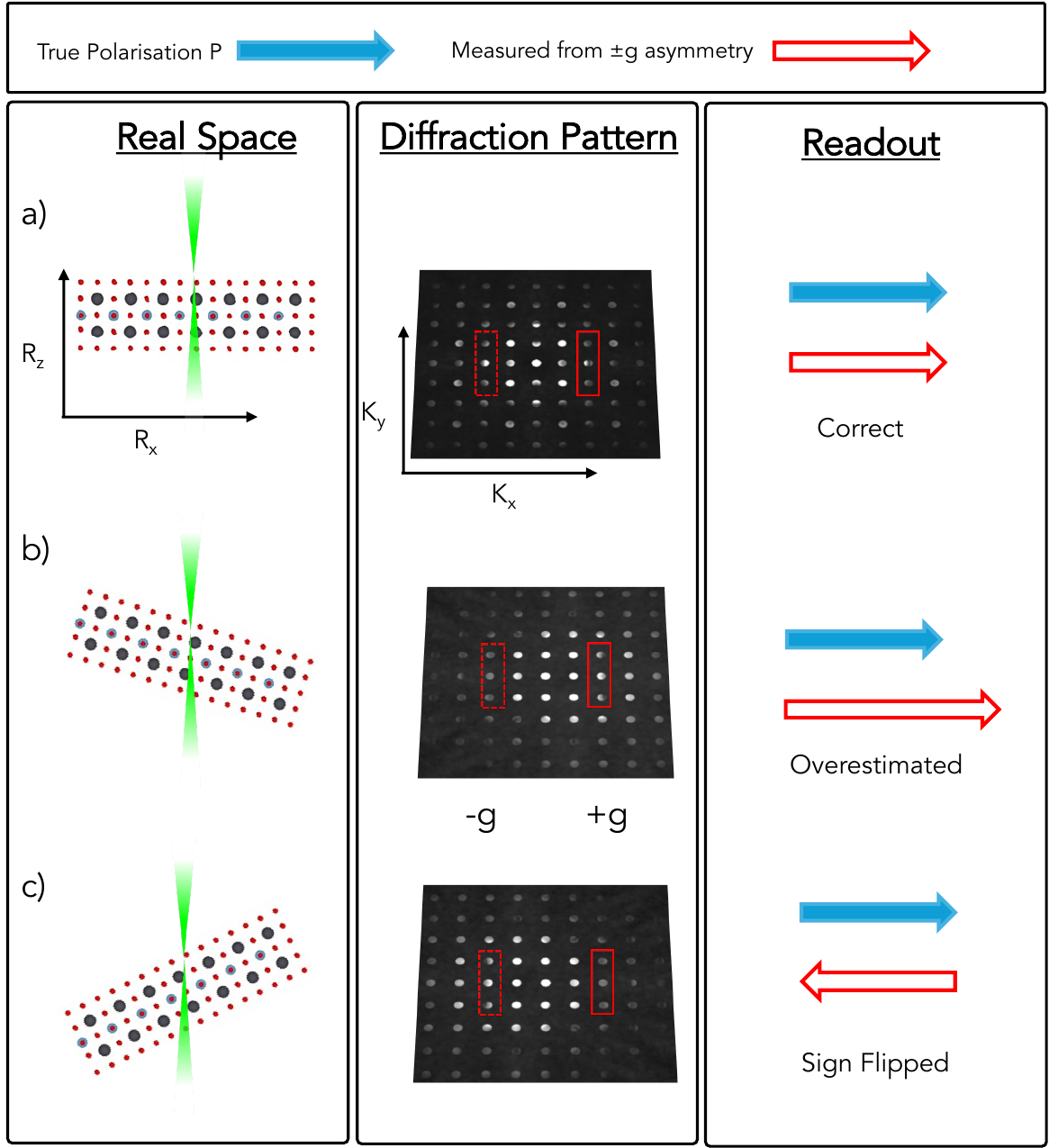}
     \caption{Illustration of the effect of sample tilt on the detected polarisation signal, in this case arising from a natural tilt of the sample. (a) Shows the on-axis reference case. In each configuration, the true polarisation direction is indicated by the solid blue arrow, and the polarisation signal detected via the Friedel disk pair method is shown by the hollow red arrow, where orange boxes highlight the regions of interest (ROIs) used for this measurement in the diffraction images. (b) Shows the case where the sample is tilted in the direction of polarisation, displaying an exaggerated calculated polarisation signal. (c) Shows the case where the sample is tilted in the opposite direction to the polarisation, with the corresponding diffraction pattern, and highlights the incorrect polarisation signal that arises in this configuration.}
    \label{fig:schematic}
\end{figure}

Even small deviations from the zone axis can significantly modify diffraction intensities and symmetry relationships within CBED patterns. Changes in Bragg disc intensity\cite{https://doi.org/10.1107/S0108767388004532, shaoEmergentChiralityPolar2023, zeltmannDisentanglingTiltPolarization2023}, distortions of Kikuchi band asymmetry\cite{https://doi.org/10.1107/S0108767388009304}, and the appearance of additional higher-order Laue zone (HOLZ) features can all produce signals that compete with, mimic, or obscure the comparatively weak signatures associated with ferroelectric polarisation\cite{jonesHigherOrderLaue1977}. As the degree of tilt increases, these effects can dominate the diffraction pattern, leading to systematic errors in polarisation quantification and, ultimately, the loss of interpretability of polarisation-related features. 

The tilt sensitivity of these signals has been noted before: Bragg-pair asymmetry in PTO reverses under mistilts of 0.04\degree\cite{shaoEmergentChiralityPolar2023} and in multilayers under 0.3\degree\cite{zeltmannDisentanglingTiltPolarization2023}, while Kikuchi-band asymmetry has been reported to tolerate 0.23 \cite{shaoEmergentChiralityPolar2023}. These observations come from single systems and geometries, and the two methods have not been compared systematically on the same patterns.

Establishing the limits within which polarisation signatures remain distinguishable from tilt-induced artefacts is therefore essential for the reliable application of diffraction-based ferroelectric characterisation methods. Such knowledge is particularly important for automated and machine-learning-assisted analysis workflows, where robust feature identification is a prerequisite for reliable operation. Quantifying the angular thresholds at which polarisation-related diffraction features become degraded or ambiguous defines the practical operating window of these approaches and provides a framework for assessing their applicability to complex ferroelectric microstructures.

In this study, PbTiO$_3$ (PTO), BaTiO$_3$(BTO), and \ce{Cu3B7O13Cl} (boracite) were investigated as representative ferroelectric systems. The effects of specimen tilt on polarisation mapping were quantified using two diffraction-based approaches: Friedel-pair intensity asymmetry and Kikuchi-band asymmetry analysis. These materials were selected to span a broad range of ferroelectric behaviours, crystal symmetries, and polarisation mechanisms. PTO serves as a model proper ferroelectric, possessing a comparatively simple tetragonal perovskite structure, a large spontaneous polarisation, and a high Curie temperature. Its strong polar distortion and well-understood behaviour make it an ideal benchmark system for evaluating the robustness of diffraction-based polarisation measurements. BTO, while structurally related to PTO, exhibits a smaller ferroelectric distortion and weaker spontaneous polarisation, providing a more demanding test of feature detectability and sensitivity to specimen misorientation. Boracite represents a distinctly different class of polar oxide material. Unlike the perovskite oxides, it possesses a more complex crystal structure in which polarisation emerges through a coupled structural distortion rather than simple cation off-centring, giving rise to fundamentally different diffraction signatures. Together, these systems provide a diverse testbed spanning strong and weak ferroelectric distortions, proper and improper ferroelectric mechanisms, and both perovskite and non-perovskite crystal chemistries.

By examining the evolution of polarisation-related diffraction features with increasing specimen tilt across these three materials, we establish quantitative limits for the reliable extraction of polarisation information and assess the extent to which these limits depend on the underlying ferroelectric mechanism and crystal structure.

\begin{figure*}
    \centering
    \includegraphics[width=0.9\linewidth]{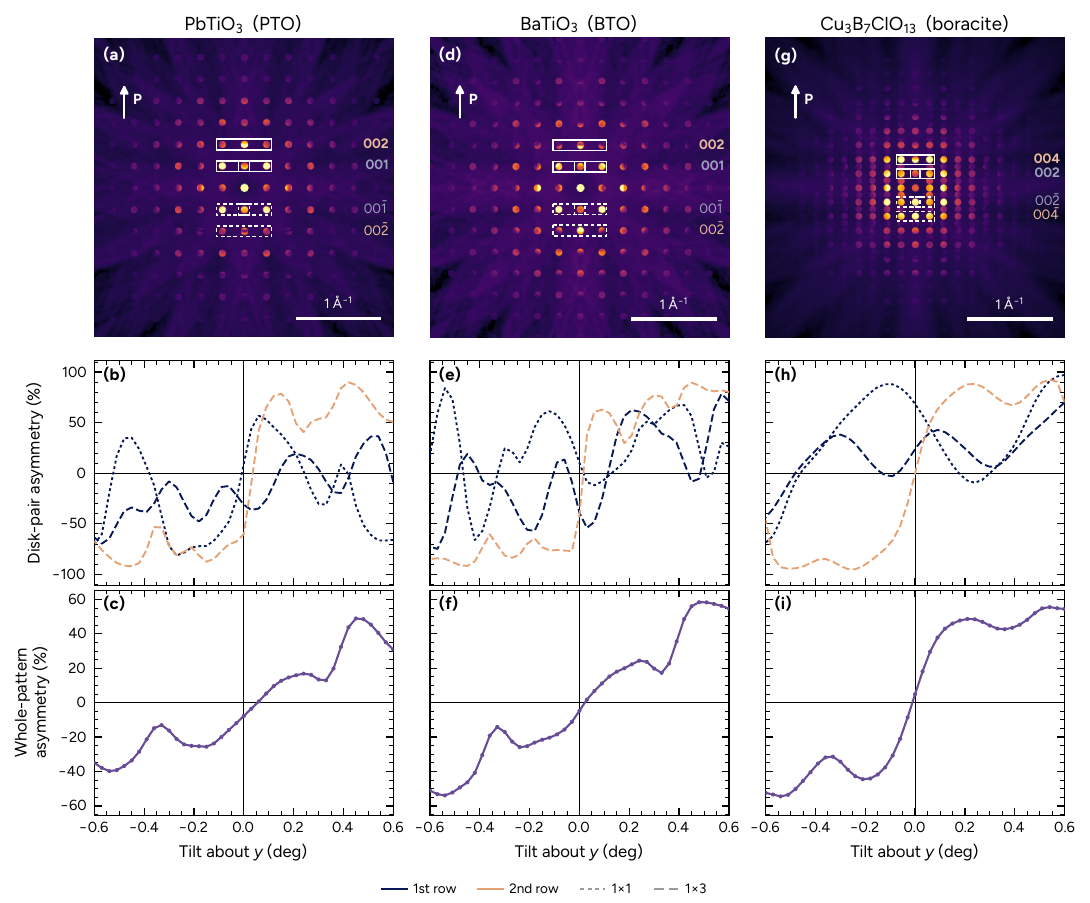}
    \caption{(a, d, g) Zero-tilt diffraction patterns of each material showing the Friedel disk pairs used.
Solid boxes are labelled with the reflection at their centre; dashed boxes enclose its Friedel partner.
Boxes enclose either one disk (1$\times$1) or three disks in a row (1$\times$3). The arrow marks the
polarisation direction. (b, e, h) Calculated intensity mismatch between each of the Friedel disk pairs at different tilting angles. The signal is positive if the ROI on the side P points to is brighter. At the simulated thickness (60nm for PTO and BTO, 48nm for boracite), the ROIs produce negative values at zero tilt for PTO and BTO but positive for boracite (see Fig.~\ref{fig:thickness-series}). (c, f, i) Whole pattern asymmetry, the same quantity as Fig.~\ref{fig:short-kikuchi}(c, f, i).}
    \label{fig:less-disks}
\end{figure*}

\section{Results}
\subsection{Friedel Disk Pair Intensity Analysis}

\begin{figure*}
    \centering
    \includegraphics[width=0.9\linewidth]{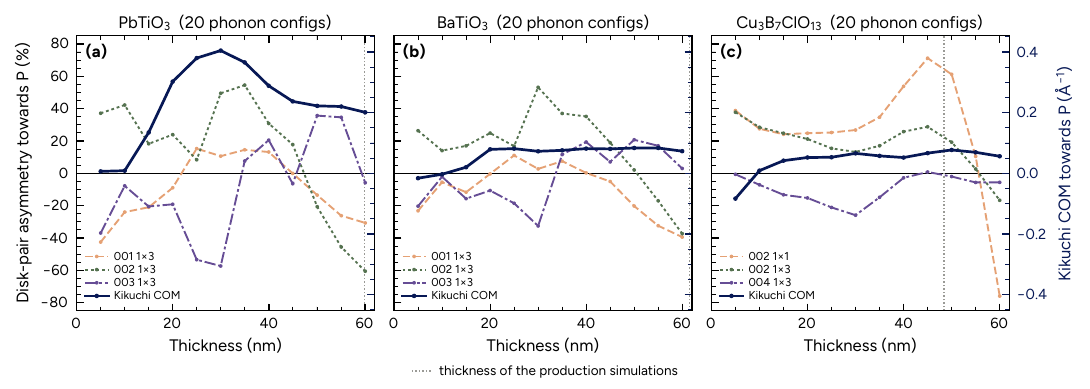}
    \caption{Zero-tilt polarisation signals as a function of specimen thickness for (a) PTO, (b) BTO and (c) boracite, from exit planes recorded every 5 nm within a single 60 nm multislice calculation (20 frozen-phonon configurations). Disk-pair asymmetries (left axis) and the Kikuchi COM (right axis) are both given relative to the polarisation direction ('towards P'). The dotted line marks the thickness of the tilt simulations. The Friedel-pair signals oscillate in sign with thickness, whereas the Kikuchi COM keeps a single sign from 15~nm in all three materials.}
    \label{fig:thickness-series}
\end{figure*}

In a ferroelectric crystal, the absence of inversion symmetry allows structure factors $F_{hkl}$ and $F_{\bar{h}\bar{k}\bar{l}}$ to differ in phase. Through dynamical electron scattering, interference between multiple diffracted beams can then produce measurable intensity differences between Friedel-related Bragg disc pairs. The sign of this asymmetry is set by the polarisation direction, but whether the disk on the side of the polarisation direction is brighter than its mirrored partner depends on the dynamical scattering conditions, in particular the specimen thickness (Fig.~\ref{fig:thickness-series})~\cite{nguyenTransferringOrbitalAngular2023b, eldredPolarizationFluctuationBaTiO32022}. Disks measured perpendicular to the polarisation direction show no asymmetry and cannot be used as a polarisation metric. Under dynamical scattering conditions, these differences are further amplified through multiple-beam interference, making the disk intensity ratio a sensitive probe of local polarisation.

However, this same sensitivity renders the technique inherently brittle with respect to beam tilt. A small tilt of the incident beam away from the zone axis (± 0.02\degree) introduces a smooth, monotonically varying intensity imbalance between Friedel disk pairs that grows approximately linearly with tilt angle in the small-angle regime. This tilt-induced asymmetry competes directly with the crystallographic signal since both manifest as an intensity difference between the same disk pairs. At larger misorientations, the relationship becomes non-linear as the diffraction condition changes substantially, and the assumption of a well-defined zone axis geometry breaks down entirely, making the polarisation signal increasingly difficult to disentangle from the tilt contribution.

To quantify polarisation direction in PTO, BTO, and boracite, intensity asymmetry was evaluated using Regions of Interest (ROI) centred on Bragg disks along the polarisation direction. The integrated intensity of each ROI was compared against that of its Friedel pair, defined as the corresponding region centred on the inversion-related Bragg disk, and the percentage intensity difference between the two was computed as the asymmetry metric (equation detailed in methods section). For the crystal symmetries and zone axes investigated here, Friedel-related Bragg disc pairs perpendicular to the polarisation direction do not provide a reliable measure of polarisation-dependent intensity asymmetry and were not considered in the calculations. In total, three ROIs were used to calculate the intensity mismatch in each diffraction pattern: an ROI spanning a single disk pair on the 001 reflection (002 for boracite), a multi-disk ROI (spanning three disks on the same reflection spanning the central disk and its two nearest neighbour disks perpendicular to the direction of polarisation) on the 001 reflection (002 for boracite) and another multi-disk ROI on the 002 reflection (004 for boracite). Their corresponding intensity asymmetry was then evaluated over 41 sample tilts from \(-\)0.6\degree to 0.6\degree to assess which disk pairs exhibited the signal most resistant to tilting effects. The best-performing ROIs are shown in Figure \ref{fig:less-disks}, and the full range of tested ROIs can be found in Figure \ref{fig:full-disks}. A positive value indicates that the ROI on the true polarisation side carries greater integrated intensity than its inversion-related counterpart (see Methods for simulation details). At the simulated thickness (60nm for PTO and BTO and 48nm for boracite), the best-performing ROIs for PTO and BTO are negative at zero tilt and positive for boracite (Sec.~\ref{sec:thickness}).

The different ROI intensity differences yielded substantially different results across the three material systems. In all samples, the ROI centred around Friedel disk pairs further away from the central beam disk produced the signal least robust to tilt in both the single-disk and multi-disk configurations, while the disk pairs closer to the central beam yielded the results most resilient to sample tilt. Even the best-performing ROI in each material reverses sign at tilts of 0.14\degree (PTO), 0.04\degree (BTO), and 0.18\degree (boracite) (Table~\ref{tab:tilt-failure}). The failure is strongly asymmetric in tilt direction: on the opposite side, PTO retains its zero-tilt sign beyond \(-\)0.6\degree, BTO to +0.114\degree and boracite to -0.460\degree.

For reference, an additional calculation was performed, where the relative summed intensity difference between the top and bottom halves of the diffraction patterns was calculated for each image at each tilting angle. The results in Figure \ref{fig:less-disks}c show that this method changes sign at 0.011-0.055\degree across all three materials, and hence shows that the Friedel disk pair method captures real signal by considering only the ROI intensity regions containing the Friedel disk pairs, rather than an overall intensity shift between the top and bottom halves of the image.

\subsection{Kikuchi Band Intensity Asymmetry Analysis}

\begin{figure*}
    \centering
    \includegraphics[width=0.9\linewidth]{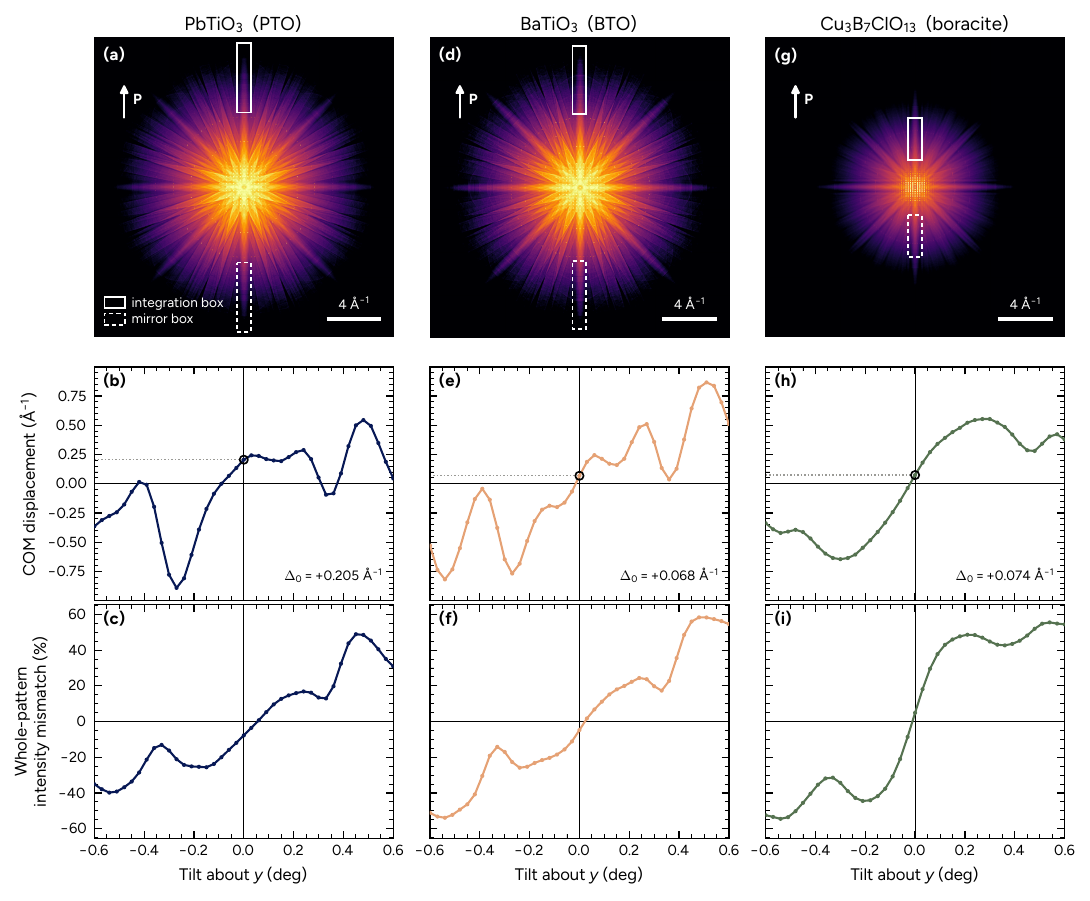}
    \caption{(a, d, g) Boxes shown on the full captured diffraction patterns of the PTO, BTO and boracite simulation images. The dashed boxes represent the mirrored boxes that are being used to calculate the centre-of-mass displacement of the Kikuchi bands. The visible circular apertures are from bandwidth limiting of the simulated wavefunction. Box angles are 108-208mrad for PTO and BTO and 40-100mrad for boracite, all ±10mrad wide (b, e, h) Centre-of-mass (COM) y displacement is calculated across the range of y tilt angles of the incident beam in simulations. \(\Delta_{0}\) indicates the value at zero-tilt. The method fails when the curve changes sign from its zero-tilt value. (c, f, i) Complete intensity mismatch between the top and bottom sections of the image. The signal is negative (opposite to P) in PTO and BTO at zero tilt and positive in boracite.}
    \label{fig:short-kikuchi}
\end{figure*}

To assess the effect of beam tilt on Kikuchi-based polarisation quantification, an analogous simulation framework to the Friedel disk intensity analysis was employed. Rather than integrating disk intensities, the full CBED pattern is retained and processed following the methodology of Shao et al. \cite{shaoEmergentChiralityPolar2023}, in which regions of interest are placed over pairs of inversion-related Kikuchi bands along the polar axis at 108-208 mrad (±10 mrad wide) for PTO and BTO and 40-100 mrad for boracite (also ±10mrad wide), whose bandwidth-limiting aperture lies at 154 mrad. The full diffraction pattern is masked to each ROI in turn, and the component of the Centre Of Mass (COM) of the combined masked intensity along the polar axis is computed. The COM displacement from the pattern centre is converted to Å\(^{-1}\) and used as a polarisation metric, with a positive value indicating that the intensity distribution is displaced towards the side of the pattern corresponding to the true polarisation direction. This metric was tracked as a function of tilt angle to establish the range over which the method remains sensitive to polarisation direction.

Unlike Bragg disk intensities, which depend on the excitation condition of individual reflections at specific points in reciprocal space, Kikuchi bands arise from diffusely scattered electrons that have been sampled from a broad region of reciprocal space through thermal scattering before being diffracted. The resulting asymmetry signal is therefore an integration over many reciprocal space points and is expected to be less sensitive to the excitation of any single reflection\cite{shaoEmergentChiralityPolar2023,maclarenOriginDifferentialPhase2015, shaoLatticeRotationVortexCharged2017}. As with the Friedel disk pair calculations, a range of tilt angles from –0.6\degree to 0.6\degree was simulated. 41 evenly spaced sample tilt orientations were sampled across this range. Placing the box on lower-order bands does not improve the tolerance: the 6--26~mrad box gives the wrong polarity at zero tilt in all three materials, and the other positions tested fail at most 0.002\degree later than the box used here, while binning the pattern to 0.1~\AA$^{-1}$ pixels or tapering the box edges changes the tolerance by at most 0.002\degree (Table~\ref{tab:kikuchi-box}).

The Kikuchi COM retains its zero-tilt sign only to 0.09\degree in PTO, 0.015\degree in BTO and 0.02\degree in boracite, compared with 0.14\degree, 0.04\degree and 0.18\degree for the best Friedel-pair ROI. However, the failure is one-sided and tilting in the opposite direction; PTO survives to +0.31\degree, BTO to beyond +0.6\degree and boracite to beyond +0.6\degree. In all three materials the Kikuchi COM points along the polarisation direction at zero tilt.

With no tilt along the polar axis (y=0) and instead tilted over the axis perpendicular to the polarisation direction (x-direction) at ±0.6\degree, PTO and boracite keep their zero-tilt sign throughout, and BTO reverses at an interpolated value of 0.42\degree. Hence, the response in tilt perpendicular to the polarisation direction has a larger tolerance range for retaining its zero-tilt sign, however still distorts in strength. These results are shown in~\ref{fig:kikuchi-grid} (Appendix).

\subsection{Zero-Tilt Polarity and Specimen Thickness}
\label{sec:thickness}

Before comparing tilt tolerances, we test whether the zero-tilt sign of each metric identifies the polarisation direction at all. Fig.~\ref{fig:thickness-series} shows the zero-tilt signals from 5 to 60~nm (see Methods). The Friedel-pair asymmetry changes sign with specimen thickness, consistent with previous reports of thickness-dependent contrast reversals arising from dynamical scattering\cite{nguyenTransferringOrbitalAngular2023b, eldredPolarizationFluctuationBaTiO32022}: in PTO, for example, the 002 1$\times$3 ROI is positive from 0 to 45~nm and negative beyond. The brightest disk in a Friedel disk pair therefore depends on the specimen thickness as well as on the polarisation. The Kikuchi COM, in contrast, points along the polarisation direction at every thickness from 15 to 60~nm in all three materials, including the improper ferroelectric boracite, whose small net polarisation follows the relative displacement of the \ce{Cu} and \ce{Cl} sublattices\cite{hardyPolarizationDomainMapping2025}.

At the thickness of the tilt simulations (60~nm for PTO and BTO, 48~nm for boracite), the two methods therefore assign opposite polarities in PTO and BTO, and agree only in boracite. The zero-tilt sign of the Kikuchi COM gives the correct polarity at all thicknesses above 15~nm, whereas the Friedel-pair-calculated polarisation direction can only be trusted once it has been checked against a multislice simulation at the specimen thickness.

\section{Discussion}
These results provide quantitative tilt sensitivity limits for the two principal polarisation analysis methods in electron diffraction: Bragg disk intensity asymmetry and Kikuchi band intensity asymmetry mapping. These were measured across three structurally and mechanistically distinct ferroelectric systems, and the mechanisms are expected to apply to other polar materials. 

A concern identified in this study is that beam tilt not only risks reversing the sign of the polarisation metric and producing an incorrect reading but also distorts the magnitude of the polarisation signal. In experimental settings where the precise tilt angle is unknown, this distortion could lead to systematic errors in both the direction and the apparent strength of the local polarisation, producing misleading quantitative maps without any obvious indicator of failure.

\begin{figure*}
    \centering
    \includegraphics[width=0.9\linewidth]{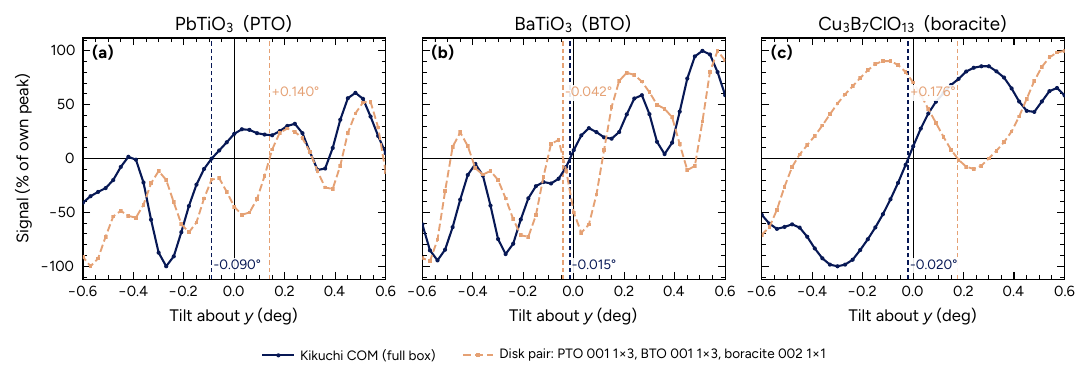}
    \caption{Comparison between the Kikuchi COM and best-performing Friedel disk pair ROI for each material over the same tilting range. The calculations from the best-performing Friedel disk ROI for each material are displayed on the comparison plot here. The y-axis is normalised against its largest absolute value over this tilt range. The point at which each method no longer holds its zero-tilt sign is also marked on the x-axis on each plot.}
    \label{fig:comparison-fig}
\end{figure*}

The Friedel-disk pair method proved highly ROI-dependent. Most ROIs reversed sign within 0.01-0.05\degree, but specific, system-dependent disk ROIs produced more robust tolerances, where the best retain their zero-tilt sign at 0.14\degree (PTO), 0.04\degree (BTO) and 0.18\degree (boracite), creating a wider window of reliability. The fundamental fragility of this technique stems from the fact that tilt-induced asymmetry and polarisation signal both manifest as an intensity mismatch between the same Friedel disk pairs and cannot be decoupled in a single-pair measurement. Fitting many reflections with a dynamical model is one route to separating them\cite{zeltmannDisentanglingTiltPolarization2023}. 

Contrary to the expectation that Kikuchi-based polarisation mapping is less sensitive to specimen mis-tilt, the Kikuchi COM method retains its zero-tilt asymmetry sign over a narrower angular range than the Friedel-disc intensity asymmetry method (0.09\degree, 0.015\degree and 0.02\degree, compared with 0.14\degree, 0.04\degree and 0.18\degree for PTO, BTO and boracite, respectively). This advantage comes with an important caveat. As shown in Sec.~\ref{sec:thickness}, the zero-tilt sign of the Friedel pairs is itself set by the specimen thickness, so the sign they retain only identifies the polarity once it has been checked against a dynamical simulation at that thickness. The Kikuchi COM gives the correct polarity without this check, but only within a few hundredths of a degree of the zone axis. These results are shown in full in Table~\ref{tab:tilt-failure}. Although the Kikuchi method is evaluated on the same patterns as the Friedel disk pair methods, the Kikuchi method's polar signal has a 1-3\% brightness imbalance between the ROI and its mirror, and a tilt of 0.1\degree changes the imbalance by 2-12\%, so the reading reverses once the Kikuchi pattern has moved by only 2-9 detector pixels. 

This ordering differs from that reported by Shao et al.\cite{shaoEmergentChiralityPolar2023}, whose simulations of PbTiO$_3$ (101~nm, 2~mrad) found Bragg-pair asymmetries reversing at 0.04\degree and the Kikuchi asymmetry tolerating 0.23\degree. However, the tilt axis and direction are not specified there. Failure of both methods varies highly on tilting direction of the sample, with tilts perpendicular to the polarisation axis being far less harmful to the sign, and the tolerance also varies with specimen thickness. A single tolerance value can therefore differ substantially depending on the tilt geometry and thickness at which it is evaluated.

\begin{table}
\small
\setlength{\tabcolsep}{4pt}
\begin{tabular}{lrrrr}
\toprule
 & $S_0$ & $t_-$ (\degree) & $t_+$ (\degree) & $\theta_\mathrm{tol}$ (\degree) \\
\midrule
\multicolumn{5}{l}{\textit{Kikuchi COM (\AA$^{-1}$)}} \\
PTO & $+0.205$ & $-0.090$ & $+0.311$ & $0.090$ \\
BTO & $+0.068$ & $-0.015$ & $>0.6$ & $0.015$ \\
Boracite & $+0.074$ & $-0.020$ & $>0.6$ & $0.020$ \\
\addlinespace
\multicolumn{5}{l}{\textit{Friedel disk pair (\%)}} \\
PTO, 001 1$\times$3 & $-31.2$ & $<-0.6$ & $+0.140$ & $0.140$ \\
BTO, 001 1$\times$3 &  $-39.3$ & $-0.042$ & $+0.114$ & $0.042$ \\
Boracite, 002 1$\times$1 & $+68.6$ & $-0.460$ & $+0.176$ & $0.176$ \\
\addlinespace
\multicolumn{5}{l}{\textit{Top/bottom halves (\%)}} \\
PTO & $-7.7$ & $<-0.6$ & $+0.055$ & $0.055$ \\
BTO & $-4.7$ & $<-0.6$ & $+0.022$ & $0.022$ \\
Boracite & $+4.9$ & $-0.011$ & $>0.6$ & $0.011$ \\
\bottomrule

\end{tabular}
\caption{Tilt tolerance of the three polarisation metrics, evaluated on the same simulated patterns (tilt along the polar axis \(y\); 0.03\degree steps over $\pm0.6\degree$). $S_0$: signal at zero tilt. $t_-$, $t_+$: first tilt on each side of zero at which the signal loses its zero-tilt sign (calculated with linear interpolation). $\theta_\mathrm{tol}$ is the smaller of the two. $>0.6$ means the sign is kept over the whole simulated range on that side. For the Friedel disk pair method, the best-performing ROI is given.}
\label{tab:tilt-failure}
\end{table}

We also find that the tilt direction matters strongly. All tolerances in Table~\ref{tab:tilt-failure} are for tilts along the polarisation axis. For tilts perpendicular to the polarisation axis, the Kikuchi COM keeps its zero-tilt sign over ±0.6\degree in PTO and boracite and to about 0.4\degree in BTO (see Fig.~\ref{fig:kikuchi-grid}) and the best first-row disk ROIs (001 1$\times$3 in PTO and BTO, 002 1$\times$1 in boracite) to at least ±0.3\degree. These results show that tilt, even orthogonal to the polarisation direction, can affect the measurement of the polarisation signal, even entirely reversing its sign in some cases.

The Kikuchi COM tolerance follows the size of the zero-tilt signal, being largest in PTO ($S_0=+0.205$~\AA$^{-1}$, 0.09\degree) and similar in BTO and boracite ($S_0\approx0.07$~\AA$^{-1}$, 0.015--0.02\degree), consistent with the stronger polar distortion of PTO. The best Friedel-pair tolerance shows no such ordering (boracite $>$ PTO $>$ BTO), and is set instead by which reflections happen to be favourably excited at the simulated thickness. Because Kikuchi bands are formed by thermal diffuse scattering, they also weaken as the specimen is cooled, so, as for thin specimens, measurements at low temperature would be expected to be more susceptible to noise.

\section{Methods}
Crystallographic information files (CIFs) for PTO and BTO were obtained from the Inorganic Crystal Structure Database \cite{zagoracRecentDevelopmentsInorganic2019} (ICSD accession numbers ICSD-61168 and ICSD-34903, respectively). The \ce{Cu3B7O13Cl} boracite structure was obtained from first-principles density functional theory calculations\cite{hardyPolarizationDomainMapping2025}.
Multislice 4D-STEM simulations were performed using the abTEM software package\cite{AbTEMCodeTransmission}. Supercell dimensions were chosen to achieve an approximate specimen thickness of 48-61nm in all cases. For the simulations (both calculation methods used the same simulated diffraction images), PTO and BTO were tiled to $30 \times 30 \times 154$ unit cells (11.8nm $\times$ 11.6nm $\times$ 59.8nm for PTO,  and 12.1nm $\times$ 11.9nm  $\times$ 61.4nm for the BTO) along the [100] zone axis, while boracite was tiled to $7 \times 7 \times 40$ unit cells (16.9nm $\times$ 16.9nm $\times$ 48.4nm) along the [110] zone axis. 40 frozen phonons were used for all simulations. Atomic vibration amplitudes were parametrised using Debye-Waller factors expressed as root-mean-square displacements. For PTO, values of 0.1068 Å, 0.0872 Å, and 0.0503 Å were used for Pb, Ti, and O, respectively, and for BTO, values of 0.0785 Å, 0.0703 Å, and 0.0656 Å were used for Ba, Ti, and O, respectively. Experimental Debye-Waller factors were not available for boracite, so a uniform displacement of 0.1 Å was applied to all atomic species. A slice thickness of 1.0 Å was used for all calculations.
The electron probe was modelled at an accelerating voltage of 300 keV with a convergence semi-angle of 0.8 mrad and no aberrations. 

Diffraction patterns were simulated out to a maximum angle of 341 mrad for PTO, 334 mrad for BTO, and 238 mrad for boracite for all simulations. For antialiasing, a bandwidth limit of 2/3 the maximum sampled angle of the electron wavefunction was applied \cite{kirklandAdvancedComputingElectron2020}. All simulations also used 4096 grid points and a 2\(\times\)2 probe scan over the full supercell was averaged for each pattern.

All results were plotted with the polarisation pointing to the solid box in the diffraction pattern, so positive values of Kikuchi COM or Friedel disk pair intensity mismatch indicate an agreement with the polarisation direction.

The simulations for the data in Fig~\ref{fig:thickness-series} were completed with the same parameters as the rest of the calculations, with the only difference being that they were calculated with a configuration of 20 frozen phonons. Exit planes were recorded every 5nm from 5 to 60nm in a single calculation. The signals were also recorded with the same ROIs as in the tilt series.

For the construction of the ROIs for the Friedel disk pair methods, the ROIs are centred on the disk lattice, ±7px (PTO, BTO) or ±9px (boracite) about each disk centre. Each ROI was compared to its exact mirror through the pattern centre and \(\frac{I_{1} - I_{2}}{I_{1} + I_{2}}\) was taken as the asymmetry. \(I_1\) was taken as the ROI box which lay in the direction of the polarisation in the diffraction pattern, so that a positive calculated value corresponds to an intensity mismatch which corresponds with the correct direction of polarisation.

Polarisation directions of PTO and BTO were taken from the simulated structures. For boracite, the direction was taken along the relative displacement of the Cu and Cl sublattices, following the established mechanism \cite{hardyPolarizationDomainMapping2025, mcquaidInjectionControlledMotion2017c}. 

\section{Conclusion}

This work establishes quantitative tilt tolerances for two widely used diffraction-based polarisation mapping approaches: Friedel-pair intensity asymmetry and Kikuchi-band asymmetry analysis. These tolerances (0.015--0.18\degree) are comparable to or smaller than the angular precision achievable through goniometer-based specimen alignment alone \cite{leeAdvancedMethodAccurate2018}. These tolerances are often smaller than the local misorientations found in real ferroelectric microstructures: the lattice rotation across a tetragonal ferroelastic domain wall can be estimated from $2\arctan(c/a)-90^\circ$, giving 0.8\degree and 0.5\degree for the \ce{PbTiO3} and \ce{BaTiO3} structures used here, respectively. The corresponding crystallographic misorientation in orthorhombic boracite is 0.3\degree for the structure investigated. Other features such as strain gradients, dislocations, and specimen bending add further variation within a single scan. Therefore, even when a specimen is aligned to the zone axis, local regions may already exceed the angular limits for reliable polarisation mapping.

Furthermore, the sign of the Friedel-pair asymmetry is not a universal indicator of polarity, and can reverse with specimen thickness without any change in polarisation direction \cite{nguyenTransferringOrbitalAngular2023b, eldredPolarizationFluctuationBaTiO32022}, so absolute polarity assignment from Bragg-disc intensity asymmetry requires dynamical simulations that account for specimen thickness and crystal orientation. The Kikuchi COM, by contrast, retained the correct polarisation sign across all three materials for the simulated thickness range above 15~nm, although it was not necessarily more tolerant to specimen tilt. These results highlight the importance of considering both specimen thickness and local crystal orientation when interpreting polarisation-dependent diffraction signals.

The key finding is that changes in diffraction intensity caused by local crystal misorientation can be mistaken for changes in ferroelectric polarisation. This is particularly important when mapping ferroelectric domain walls, defects and complex polar textures, where lattice rotations are often part of the structure being investigated. Our results show that neither Friedel-pair nor Kikuchi-band analysis can be assumed to be insensitive to tilt, and that the more reliable approach depends on the material and diffraction conditions. By establishing the angular limits for both methods, this work provides a practical basis for assessing whether an observed polarisation contrast is reliable or whether local crystal orientation and specimen thickness must first be accounted for. As automated analysis pipelines and AI-assisted approaches become increasingly prevalent in ferroelectric characterisation, these tilt-dependent limitations need to be accounted for, otherwise changes in diffraction intensity caused by local misorientation may be incorrectly interpreted as genuine polarisation variations.

\section{Declarations}
\subsection{Data \& Code Availability}
All scripts and CIF files used in this study are available at \url{https://github.com/FinHardy/polar-tilt-sim}. The simulated 4D-STEM datasets supporting this study are available from the Imperial College London Helix repository at \url{https://doi.org/10.82186/aayj8-xj284}.

\subsection{Acknowledgments}
This work was made possible by the EPSRC Cryo-Enabled Multi-microscopy for Nanoscale Analysis in the Engineering and Physical Sciences EP/V007661/1. F. H., M.P.,  and M.S.C. acknowledge funding from the Royal Society Tata University Research Fellowship (URF\textbackslash R1\textbackslash 201318) and Royal Society Enhancement Award RF\textbackslash ERE\textbackslash 210200EM1. N.S. acknowledges funding from the European Union’s Horizon Europe research and innovation programme under grant agreement No. 101282745. M.S.C. acknowledges funding that supported this work from ERC CoG DISCO grant 101171966. G.T. and M.S.C. acknowledge funding from EPSRC grant EP/V001914/1 and Engineering and Physical Sciences Research Council (EPSRC) through the Materials Engineering for Advanced Devices (MEAD) Programme Grant (APB86538). G.T. acknowledges funding from EPSRC Centre for Doctoral Training in the Advanced Characterisation of Materials (CDTACM)(EP/S023259/1) and Cameca Ltd. for co-funding their PhD. We acknowledge the EuroHPC Joint Undertaking for awarding this project access to the EuroHPC supercomputer LEONARDO, hosted by CINECA (Italy) and the LEONARDO consortium through a EuroHPC Extreme Access call.


\bibliography{main}
\clearpage
\onecolumngrid
\section*{Appendix}
\setcounter{figure}{0}
\renewcommand{\thefigure}{\arabic{figure}A}
\renewcommand{\thetable}{\Roman{table}A}   

\begin{table*}[!h]

\small
\setlength{\tabcolsep}{4pt}
\begin{tabular}{lrrrrrr}
\toprule
 & \multicolumn{2}{c}{PTO} & \multicolumn{2}{c}{BTO} & \multicolumn{2}{c}{Boracite} \\
\cmidrule(lr){2-3}\cmidrule(lr){4-5}\cmidrule(lr){6-7}
Box & $S_0$ (\AA$^{-1}$) & $\theta_\mathrm{tol}$ (\degree) & $S_0$ (\AA$^{-1}$) & $\theta_\mathrm{tol}$ (\degree) & $S_0$ (\AA$^{-1}$) & $\theta_\mathrm{tol}$ (\degree) \\
\midrule
Paper box$^{a}$ & $+0.205$ & $0.090$ & $+0.068$ & $0.015$ & $+0.074$ & $0.020$ \\
\addlinespace
6--26 mrad & $-0.188$ & $0.037$ & $-0.122$ & $0.016$ & $-0.004$ & $0.001$ \\
26--55 mrad & $+0.040$ & $0.021$ & $+0.007$ & $0.002$ & $+0.048$ & $0.022$ \\
55--108 mrad & $+0.128$ & $0.065$ & $+0.051$ & $0.012$ & $+0.076$ & $0.017$ \\
\addlinespace
Paper box, 0.1~\AA$^{-1}$ pixels & $+0.204$ & $0.089$ & $+0.068$ & $0.015$ & $+0.073$ & $0.020$ \\
Paper box, tapered edges & $+0.215$ & $0.090$ & $+0.070$ & $0.015$ & $+0.074$ & $0.018$ \\
\bottomrule
\end{tabular}
\caption{Effect of Kikuchi box placement and detector sampling on the Kikuchi COM, using the same simulated patterns as Table~\ref{tab:tilt-failure} (tilt along the polar axis, 0.03\degree steps over $\pm0.6^\circ$). Box ranges are radial distances from the pattern centre along the polar axis (300~kV); each box is compared with its mirror through the centre, and its width perpendicular to the polar axis is unchanged. $S_0$: Kikuchi COM at zero tilt, positive when it points along the polarisation. $\theta_\mathrm{tol}$: smallest tilt at which the COM loses its zero-tilt sign. The binned row averages the pattern to 0.1~\AA$^{-1}$ pixels; the tapered row applies a Hann taper over half the box height. A dash marks a box range already covered by the boracite paper box (40--100~mrad).}
\label{tab:kikuchi-box}
\smallskip
{\footnotesize $^{a}$108--208~mrad for PTO and BTO; 40--100~mrad for boracite.}
\end{table*}

\setcounter{figure}{0}
\renewcommand{\thefigure}{\arabic{figure}A}

\begin{figure*}
    \centering
    \includegraphics[width=0.9\linewidth]{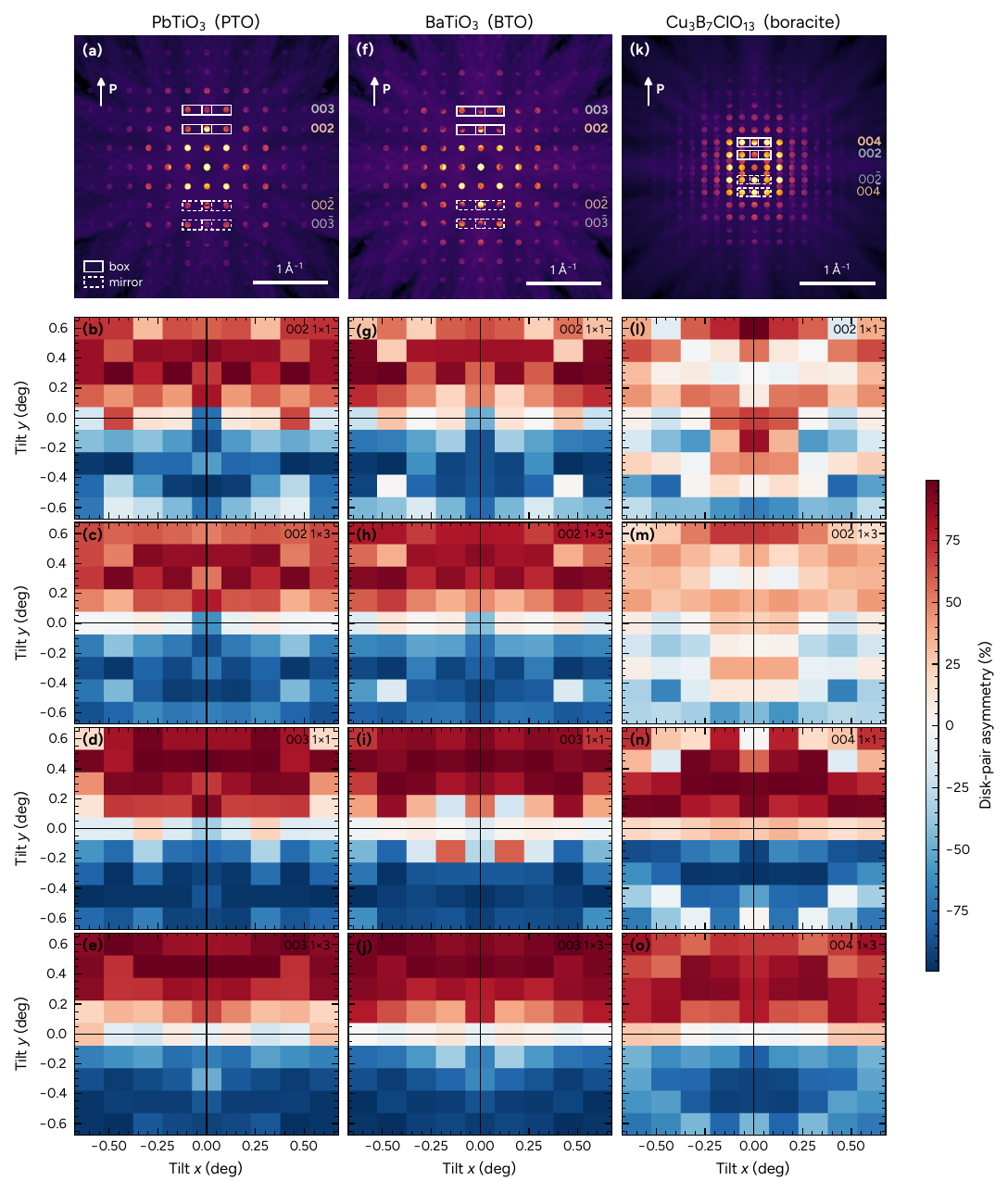}
    \caption{Tilt varied in x and y directions for the disk asymmetry analysis method with various ROIs with (a, b, c, d, e) for PTO, (f, g, h, u, i) for BTO, and (k, l, m, n, o) for Boracite.}
    \label{fig:disks-grid}
\end{figure*}

\begin{figure*}
    \centering
    \includegraphics[width=0.9\linewidth]{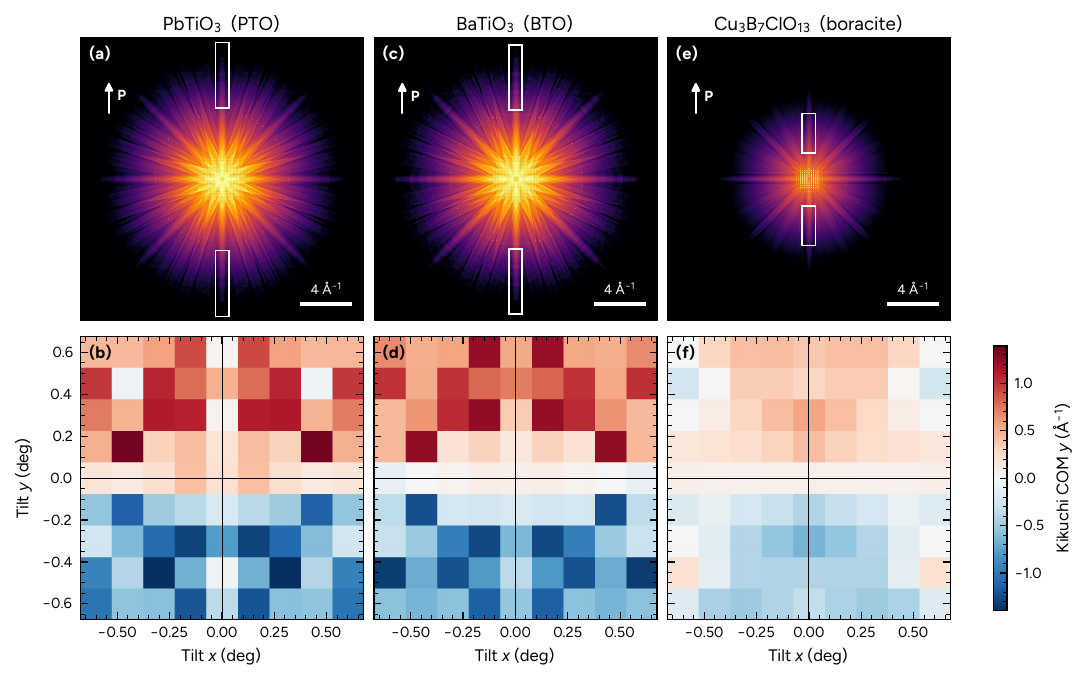}
    \caption{Tilt varied in x and y directions for the Kikuchi band asymmetry analysis method. (a, c, e) show the diffraction patterns of each material. (b, d, f) shows how the Kikuchi COM value changes as a function of tilt in the polarisation axis and plane normal to the polarisation axis.}
    \label{fig:kikuchi-grid}
\end{figure*}

\begin{figure*}
    \centering
    \includegraphics[width=0.9\linewidth]{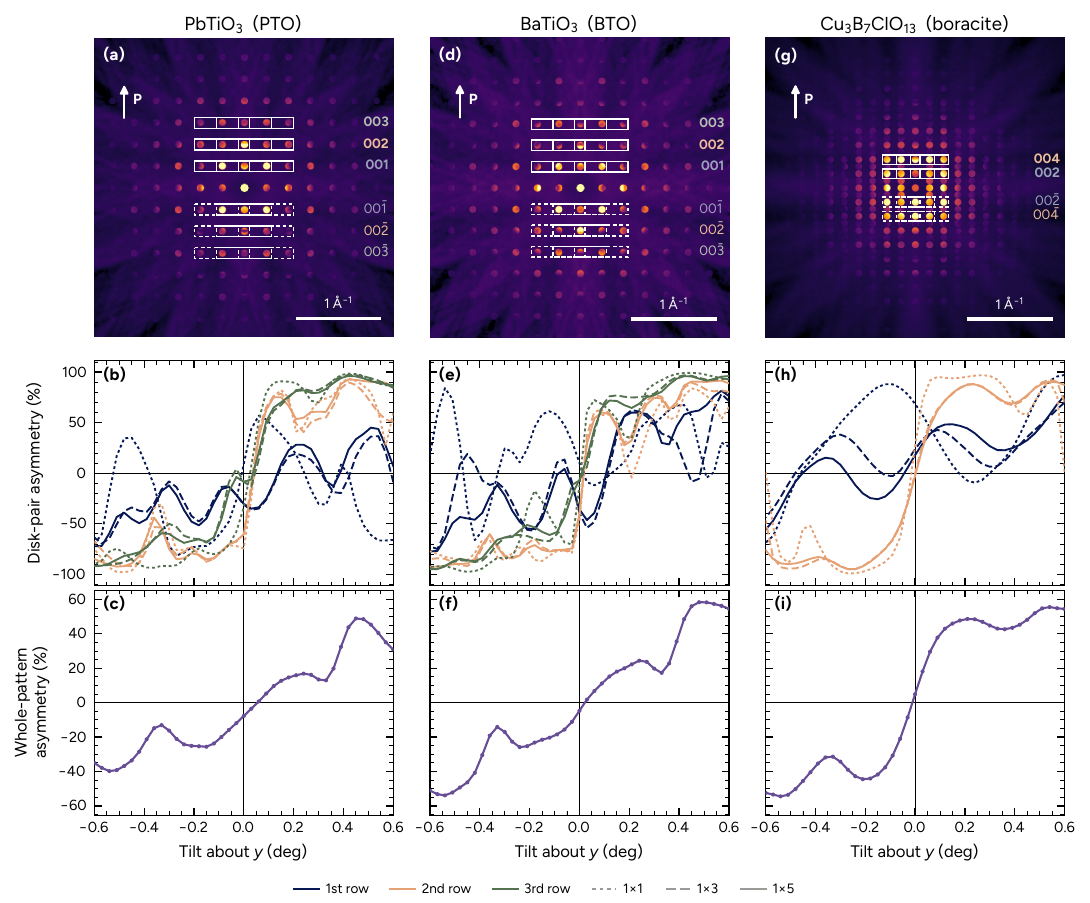}
    \caption{Calculated polarisation signals from the Friedel disk pair asymmetry method with the full range of tested ROI regions for the different Friedel disk pairs. Also showing the whole-pattern (top/bottom halves) mismatch in the bottom row.}
    \label{fig:full-disks}
\end{figure*}

\begin{figure*}
    \centering
    \includegraphics[width=0.9\linewidth]{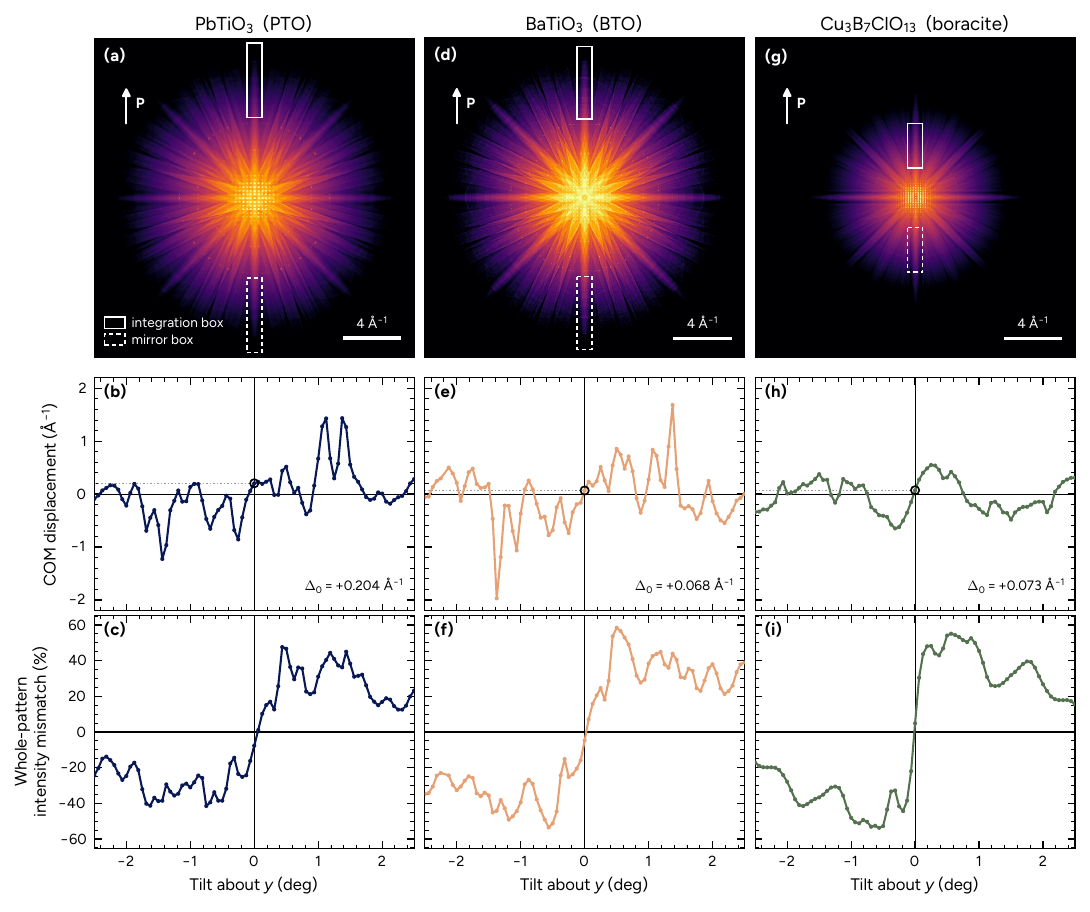}
    \caption{Calculated polarisation signals using the Kikuchi band asymmetry method at wider tilting angles from -2.5\degree to 2.5\degree (step size 0.0625\degree). Also showing the whole-pattern (top/bottom halves) mismatch in the bottom row.}
    \label{fig:long-kikuchi}
\end{figure*}

\end{document}